\documentstyle[12pt,epsfig]{article}
\begin{document}
\title{\bf Synthesizing Deuterium in Incipient Pop II Stars}
\author{Sanjay K. Pandey \footnote{Department of Mathematics, 
L.B.S.P.G. College, Gonda-271001, India}\ \  \& Daksh Lohiya 
\footnote{Department of Physics and Astrophysics,\ University of Delhi, Delhi-110007, India} \\
  Inter University Center for Astronomy and Astrophysics\\
  Ganeskhind, Pune, India\\
 email: dlohiya@iucaa.ernet.in	}
\date{}

\maketitle      
\begin{center}
Abstract
\end{center}
Sites for incipient low metallicity (Pop II) star formation
can support environments conducive to 
Deuterium production up to levels observed in the universe. This could 
have a deep impact on a ``Standard Cosmology''.

\section{Introduction:}
	Early universe (standard big-bang) nucleosynthesis [SBBN] is
regarded as a major success for the 
Standard Big Bang [SBB] Model. The results look rather good indeed. The 
observed light element abundances are taken to severely constrain cosmological 
and particle physics parameters. Deuterium, in particular, is regarded as
an ideal ``baryometer'' for determining the baryon content of the universe
\cite{olive}. 
 This follows from the fact that deuterium is burned away whenever
it is cycled through stars, and a belief, that there are no astrophysical
cites (other than SBBN), capable of producing it in its 
observed abundance \cite{eps}.
The  purpose of this  article is to admit caution in adhering to this belief.

	What would be the point of such an exercise ? 

Indeed, at the outset, drastic variations from SBBN may sound 
preposterous at this time. Confidence in SBBN stems primarily 
from $D$, $^7Li$ and $^4He$ measurements.  $D$ abundance is measured  in 
the solar wind, in interstellar clouds  and, more recently, in the 
inter-galactic medium \cite{geiss,hogan}. The belief  that no 
realistic astrophysical process other than 
the Big Bang  can  produce sufficient $D$,  lends 
support to its primordial origin. Further, 
$^7Li$ measurement [$^7Li/H \sim 10^{-10}$] in Pop II stars 
\cite{spite}  and the consensus \cite{yang} over the primordial 
value for the $^4He$ ratio $Y_p \ge 23.4 \%$ 
(by mass) suggest that
light element abundances are consistent with SBBN over nine orders of
magnitude. This is  achieved by adjusting just
one parameter,  the baryon entropy
ratio $\eta$. Alternative mechanisms for $^7Li$ production that are 
accompanied by a co-production of $^6Li$ with a later depletion of $^7Li$ 
have fallen out of favour. The debate on depletion
of $^7Li$ has been put to rest by the observation of $^6Li$ in a Pop II 
star \cite{smith}. Any depletion of $^7Li$ would have to be accompanied 
by a  complete destruction of the much more fragile $^6Li$. Within the 
SBBN scenario therefore, one seeks to account for the abundances of 
$^4He$, $D$, $^3He$ and $^7Li$ cosmologically, while $Be$, $B$ and
$^6Li$ are generated  by spallation processes \cite{olive1}.

	These results, however, do meet with occasional skepticism. 
Observation of $^6Li$, for example, requires unreasonable 
suppression of astrophysical destruction of $^7Li$. On the 
other hand, the production of $^6Li$ would be accompanied by a 
simultaneous production of $^7Li$ comparable to observed levels \cite{rees}.
This raises doubts about using observed $^7Li$ levels as a benchmark to
evaluate SBBN. 

Further the best value of $^4He$ mass fraction, 
 statistically averaged and extrapolated
to zero heavy element abundances, 
hovers around $.216 \pm .006$ for Pop II objects \cite{rana}.
Such low $^4He$ levels have also been reported     
in several metal poor HII galaxies \cite{pagel}. For example for  SBS 0335-052 
the reported value is 
$Y_p = 0.21 \pm 0.01$ \cite{ter}.
Such small values for $^4He$ would not
lead to any concordant value for $\eta$ consistent with bounds on
$^7Li$ and $D$. Of course, one could still explore a multi-parameter
non-minimal  SBBN instead of  the minimal model that just uses $\eta$
for a single parameter fit. Non-vanishing neutrino chemical potentials 
have been proposed to be ``natural'' parameters for such a venture.  
These conclusions have been criticized by
\cite{yang,ter} who rely on statistical over-emphasis on a few metal-poor
objects with a high enough $^4He$ abundance to save minimal
SBBN. On the other hand, there are 
objects reported with abysmally low $^4He$ levels. This is alarming 
for  minimal SBBN. For example, levels of $^4He$ 
inferred for $\mu$ Cassiopeiae A  \cite{ter} and from the emission
lines of several quasars \cite{peim} are as low as 5\% and 10 - 15\%
respectively. Such low levels would most definitely rule out SBBN.
At present one excludes such objects from SBBN considerations on 
grounds of ``our lack of understanding''of the environments local to these
objects. As a matter
of fact,  one has to resort to specially contrived explanations to
account for low $Y_p$ values in quasars. Considering that a host of
mechanisms for light element synthesis are discarded on grounds
of requirement of special ``unnatural'' circumstances \cite{eps}, 
it does not augur to have to resort to special explanations to contend 
with low $^4He$ emission spectra. This comment ought to be considered 
in the light of much emphasis that is 
laid on emission lines from nebulae with low metal content \cite{yang}. 
Quasars most certainly qualify for such candidates. Instead,
one merely seems to  concentrate on classes of Pop II objects and HII galaxies
that would oblige SBBN. Until the dependence of light 
element abundance on sample and statistics is gotten rid  of and / or
fully understood, one must not close one's eyes to alternatives.

	We end our overview of the status of SBBN  with a few comments.
Firstly the low metallicity that one sees in type II stars and interstellar
clouds poses a problem in SBBN. There is
no object in the universe that has quite the abundance [metallicity]
of heavier elements as is produced 
in SBB. One relies on some kind of
re-processing, much later in the history of the universe, 
to get the low observed  
metallicity in,  for example,   old clusters and 
inter-stellar clouds. This could be in the form of a generation 
of very short-lived type III stars. Such a generation of stars
may also be necessary to ionize the intergalactic medium. 
The extrapolation of $^4He$ 
abundance in type II objects and low metal (HII) galaxies, 
to its zero heavy metal abundance 
limit, presupposes that reprocessing and production of heavy
elements in type III stars is not accompanied by a significant 
change in the $^4He$ levels. A violation of this assumption,
i.e. a minute increase in $^4He$ during reprocessing (even as low
as 1 - 2 \%) would rule out the minimal SBBN. As a matter of fact, 
it is possible to 
account for the entire pre-galactic $^4He$ by such objects \cite{wagoner}.

Finally, of late \cite{lem}, the need for a careful scrutiny and a 
possible revision of the status of SBBN has also been suggested 
from  the reported
high abundance of $D$ in several $Ly_\alpha$ systems. It may be difficult 
to accommodate such high abundances
within the minimal SBBN. Though the 
status of these observations is still a matter of debate, and  
(assuming their confirmation) attempts to
reconcile the cosmological abundance of deuterium and the number of 
neutrino generations within the framework of SBB are still on, 
a reconsideration of alternate routes to deuterium presented below 
could well be worth the effort. This is specially in consideration
of the stranglehold that Deuterium has on SBB in constraining the 
baryon density upper limit to not more than some 3 to 4 \% . This
constraint has been used in SBB to make out a strong case for non -
baryonic dark matter to make up the mass estimates at galactic and
cluster scales. Relying on Deuterium that is so local environment
sensitive, to predict the nature of CDM runs the risk of ``building
a colossus on a few feet of clay''\cite{borner}

{\bf Deuterium Production:}

 To get the observed abundances of light elements besides $^4He$, 
we recall spallation mechanisms that were explored in the pre - 1976 days 
\cite{eps}. Deuterium can indeed be produced by the following spallation reactions:
$$
p + ^4He \longrightarrow D + ^3He; ~~ 2p \longrightarrow D + \pi^+;
$$
$$
2p \longrightarrow 2p + \pi^o,~ \pi^o \longrightarrow 2\gamma,~
\gamma +^4He \longrightarrow 2D.
$$There is no problem in producing Deuterium all the way 
to observed levels. The trouble is that under most conditions 
there is a concomitant over - production of $Li$ nuclei and $\gamma$ rays
at unacceptable levels. Any later destruction of lithium in turn completely 
destroys $D$. As described in \cite{eps}, figure (1) exhibits relative 
production of $^7Li$  and $D$ by spallation. It is apparent that the 
production of these nuclei to observed levels and without a collateral 
gamma ray flux is possible only if the incident (cosmic ray or any other) beam 
is energized to an almost mono energetic value of around 600 MeV. A model 
that requires nearly mono energetic particles would be rightly considered 
$ad~hoc$ and would be hard to physically justify.

     However, lithium production occurs by spallation of protons over 
heavy nuclei as well as spallation of helium over helium:
$$
p,\alpha ~+ ~C,N,O \longrightarrow Li~+~X;~~ 
p,\alpha ~+ ~Mg,Si,Fe \longrightarrow Li~+~X;~~
$$
$$
2\alpha \longrightarrow ^7Li ~+~p; ~~ \alpha ~+~D \longrightarrow p ~+~^6Li;
$$
$$
^7Be + \gamma \longrightarrow p + ^6Li; ~~ ^9Be + p \longrightarrow
\alpha +^6Li.
$$
The absence or deficiency 
of heavy nuclei in a target cloud and deficiency
of alpha particles in the incident beam would clearly suppress lithium
production. Such conditions could well be imagined in the environments
of incipient Pop II stars. 

Essential aspects of evolution of a collapsing cloud to form a low mass 
Pop II star is believed to be fairly well understood 
\cite{feig,hart}. The formation
and early evolution of such stars can be discussed in terms of
gravitational and hydrodynamical processes. A protostar would emerge from the
collapse of a molecular cloud core and would be surrounded by high angular 
momentum material forming a circumstellar accretion disk with bipolar outflows.
Such a star contracts slowly while the magnetic fields play a very important 
role in regulating collapse of the accretion disk and transferring the disk
orbital angular motion to collimated outflows. A substantial fraction of
the accreting matter is ejected out to contribute to the inter - stellar
 medium.

Empirical studies of star forming regions over the last twenty years have now
provided direct and ample evidence for MeV particle produced within 
protostellar and T Tauri systems \cite{Terekhov,Torsti}. 
The source of such accelerated 
particle beaming is understood to be violent magnetohydrodynamic (MHD)
reconnection events. These are analogous to solar magnetic 
flaring but elevated by factors of $10^1$ to $10^6$ above levels seen 
on the contemporary sun besides being up to some 100 times more frequent. 
Accounting for characteristics in the meteoritic 
record of solar nebula from integrated effects of particle irradiation
of the incipient sun's flaring has assumed the status of an industry.
Protons are the primary component of particles beaming out from the sun in 
gradual flares while $^4He$ are suppressed by factors of ten in rapid flares to
factors of a hundred in gradual flares\cite{Terekhov,Torsti}. Models of young sun
visualizes it as a much larger protostar with a cooler surface temperature
and with a very highly elevated level of magnetic activity in comparison
to the contemporary sun. It is reasonable to 
suppose that magnetic reconnection events would lead to abundant release
of MeV nuclei and strong shocks that propagate into the circumstellar
matter. Considerable evidence for such processes in the early solar nebula 
has been found in the meteoric record.
It would be fair to say that the hydrodynamical paradigms for 
understanding the earliest stages of stellar evolution is still not 
complete. However, it seems reasonable to conjecture that several features 
of collapse of a central core and its subsequent growth from acreting material
would hold for low metallicity Pop II stars. Strong magnetic fields may
well provide for a link between a central star, its circumstellar envelope and 
the acreting disk. Acceleration of jets of charged particles from the surface
of such stars could well have suppressed levels of $^4He$. Such a suppression 
could be naturally expected if the particles are picked up from an environment
cool enough to suppress ionized $^4He$ in comparison to ionized hydrogen.
Ionized helium to hydrogen number ratio in a cool sunspot temperature of 
$\approx 3000~K$ can be calculated  by the Saha's ionization formula and the 
ionization energies of helium and hydrogen. This turns out to be 
$\approx~ exp(-40)$ and increases rapidly with temperature. Any electrodynamic
process that accelerates charged particles from such a cool environment
would yield a beam deficient in alpha particles. With $^4He$
content in an accelerated particle beam suppressed in the incident beam and 
with the incipient cloud forming a Pop II star having low metallicity in the 
target, the ``no - go'' concern of (Epstein et.al. \cite{eps}) 
is effectively circumvented.
The ``no-go'' used $Y_\alpha /Y_p \approx .07$ in both the energetic particle
flux as well as the ambient medium besides the canonical solar heavy element
mass fraction. Incipient Pop II environments may typically have heavy 
element fraction suppressed by more than five orders of magnitude
while, as described above, magnetic field acceleration could accelerate beams 
of particles deficient in $^4He$.

One can thus have a broad energy band - all the way from a few MeV 
up to some 500 MeV per 
nucleon as described in the Figure 1, in which acceptable levels of deuterium
could be ``naturally'' produced. The higher energy end of the band may also
not be an impediment. There are several astrophysical processes associated with
gamma ray bursts that could produce $D$ at high beam energies with the surplus
gamma ray flux a natural by product.

{\bf Conclusions:}

Our understanding of star formation has considerably evolved since 1976. 
SBBN constraints need to be reconsidered in view of empirical evidence
from young star forming regions. These models clearly imply that 
spallation mechanism can lead to viable and natural production of 
Deuterium and Lithium in the incipient environment of Pop II stars.
One can conceive of a cosmological model in which 
early universe nucleosynthesis produces the desired primordial levels
of $^4He$ but virtually no $D$. Such a situation can arise in SBBN itself 
with a high baryon entropy ratio $\eta$. In such a universe, in principle,
Deuterium and Lithium  can be synthesized up to acceptable levels 
in the environment of incipient Pop II stars.

In SBB, hardly any metallicity is produced in the very early universe. 
Metal enrichment is supposed to be facilitated by a generation of 
Pop III stars. Pop III star formation from a pristine material is not 
well understood till date in spite of a lot of effort that has been 
expanded to that effect in the recent past \cite{sneider}. It is believed
that with metallicity below a critical transition metallicity 
($Z_{cr} \approx 10^{-4} Z_\odot$), masses of Pop III stars would be 
biased towards very high masses. Metal content higher than $Z_{cr}$ 
facilitates cooling and a formation of lower mass Pop II stars. In SBB,
the route to Deuterium by spallation discussed in this article would have 
to follow a low metal contamination by a generation of Pop III stars.

Deuterium production by spallation discussed in this article would be good
news for a host of slowly evolving cosmological models \cite{kapl,annu}.
An FRW model with a linearly evolving scale factor enjoys concordance with
constraints on age of the universe and with the Hubble data on SNe1A.
Such a linear coasting is consistent with the right amount of helium
observed in the universe and metallicity yields close to the lowest 
observed metallicities. The only problem that one has to contend with is 
the significantly low yields of deuterium in such a cosmology. In such 
a model, the first generation of stars would be the low mass Pop II stars
and the above analysis would facilitate the desired deuterium yields.

In SBB, large-scale production and recycling of metals through exploding early
generation Pop III stars leads to verifiable observational 
constraints. Such stars would be visible as 27 - 29 magnitude stars
appearing any time in every square arc-minute of the sky.
Serious doubts have been expressed on the existence and
detection of such signals \cite{escude}. The linear coasting cosmology would do 
away with the requirement of such Pop III stars altogether.

\vfil\eject  
\begin{center}
\begin{figure}
\epsfig{file=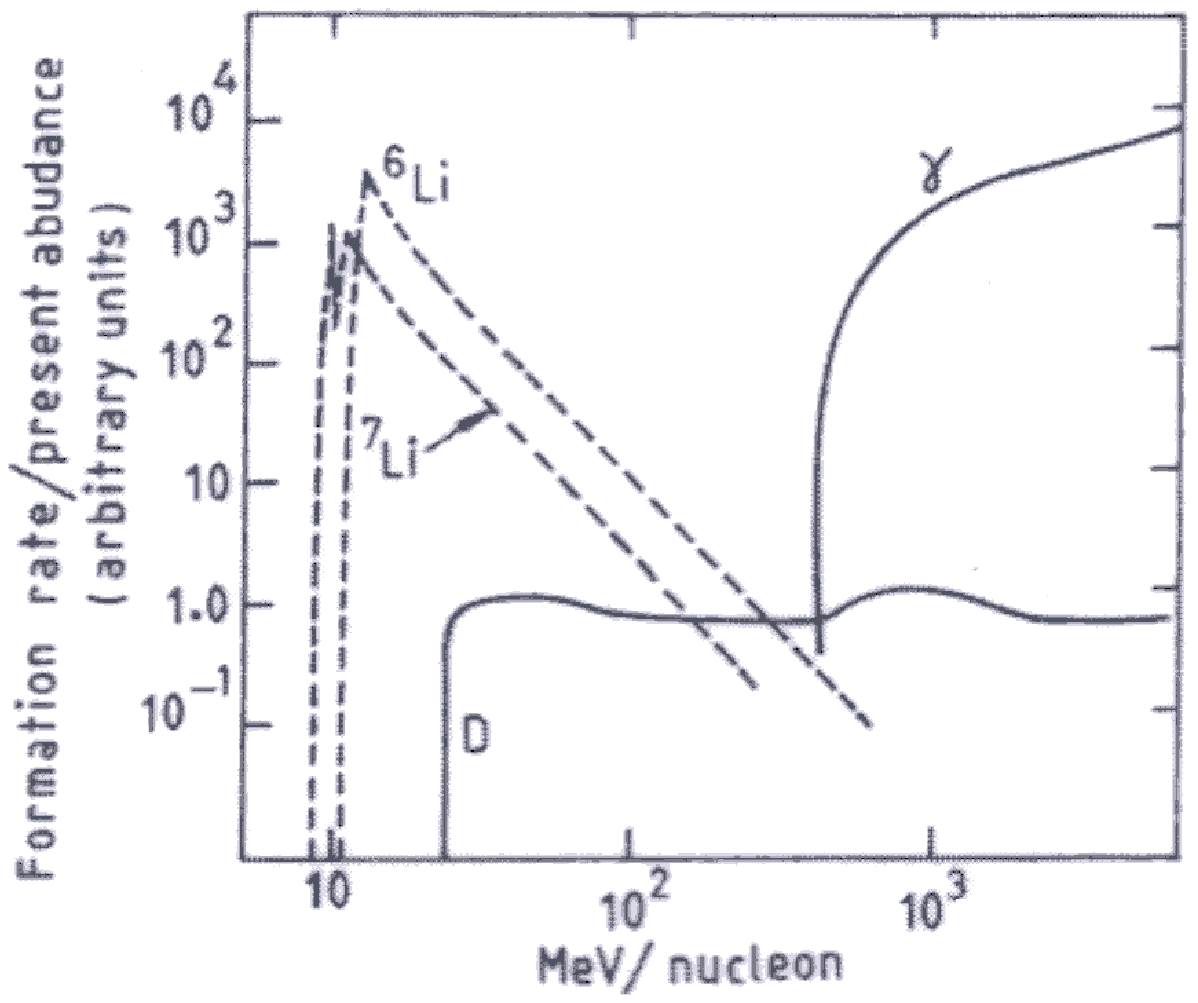,height= 5.0cm, width=15.0cm}
\label{fig1}
\end{figure}
\end{center}
\vskip 1cm

\vfil\eject
{ Acknowledgment.} \\ The authors acknowledge IUCAA for support under 
the IUCAA Associate Program. \\
\vspace{0.5cm}

\bibliography{plain}

\end{document}